%% file: aa14174-10.tex
\documentclass[structabstract]{aa}
\usepackage{graphicx}
\usepackage{txfonts}
\begin{document}
    \input{aa14174_abstract}

    \maketitle

    \input{aa14174_pulp}
    \input{aa14174_appendix}

%
\bibliographystyle{aa}      
\bibliography{Xtinct}    
%
%

\end{document}

%% file: aa14174_abstract.tex
   \title{The extinction law for molecular clouds}

   \subtitle{Case study of B\,335}

  \author{S. Olofsson
          \inst{1}
          \and
          G. Olofsson\inst{1}\fnmsep\thanks{Based on observations collected
          at the European Southern Observatory, Chile (ESO programme 077.C-0524(A))}
       }

   \institute{Stockholm Observatory, Stockholm University, Astronomy Department
              AlbaNova Research Centre, SE-106 91 Stockholm\
              \email{sven@astro.su.se}
    }
   \date{Received  2 February 2010; accepted 9 June 2010}


  \abstract
   {The large optical and near-IR surveys have made it possible to investigate
   the properties of dark clouds by means of extinction estimates.
   There is, however, a need for case studies in more detail in order to investigate
   the basic assumptions when, say, interpreting reddening in terms of column density.}
   {We determine the extinction curve from the UV to the near-IR
   for molecular clouds and investigate
   whether current models can adequately explain this wavelength dependence of the extinction.
   The aim is also to interpret the extinction in terms of $H_{2}$ column density. }
   {We  applied five different methods including
   a new method for simultaneously determining
   the reddening law and the classification of the background stars.
   Our method is based on multicolour observations and a grid of model atmospheres.}
   {We confirm that the extinction law can be adequately described by a single parameter,
   $R_{V}$ (the selective to absolute extinction), in accordance with earlier findings.
   The $R_{V}$ value for B\,335 is $R_{V}\,=\, 4.8$.
   The reddening curve can  be accurately reproduced by model calculations.
   By assuming that all the silicon is bound in silicate grains,
   we can interpret the reddening in terms of column density,
   $N_{H}\,=\,4.4\ (\pm 0.5)\cdot\,10^{21}\,E_{I-K_{s}}\ cm^{-2}$, corresponding to
   $N_{H}\,=\,2.3\ (\pm 0.2)\cdot\,10^{21}\cdot A_{V}\ cm^{-2}$, close to that of the diffuse ISM,
   $(1.8\ - 2.2)\cdot 10^{21}\ cm^{-2}$.

   We show that the density of the B\,335 globule outer shells
   can be modelled as an evolved Ebert-Bonnor gas sphere with $\rho\,\propto r^{-2}$,
   and estimate the mass of this globule to $2.5\,M_{\odot}$}
   {}

   \keywords{B335 --
             Bok globule --
             clouds --
             dust --
             visual extinction
             }

%% file: aa14174_pulp.tex
\section{Introduction}
The {\it extinction curve},
i.e. the wavelength dependence of the interstellar extinction due to dust particles,
has been the subject of numerous investigations in the past fifty years (as exemplified in Table \ref{description}).
The main driver for these efforts has been the need for restoring
the intrinsic spectral properties of the targets,
but the extinction curve also carries information on the properties of the dust particles
and their origin and development. For a recent review, see \cite{2003astro.ph..4488D}.
The extinction curve has been defined well for the diffuse ISM (interstellar matter),
and even though variations have frequently been reported for different lines of sight,
it has been found that these variations can be described
by a functional form with only one parameter, $R_{V}\,=\,A_{V}/E_{B-V}$
as shown by  \cite{1989ApJ...345..245C} (hereafter \emph{CCM}),                                           
who also found that the extinction in a few lines of sight
through the outskirts of dark clouds can be defined in the same way.
However, as the purpose of their investigation was to include the strongly variable 217.5 nm bump,
the list of background stars is restricted to O and
early B stars for which the emission from the surrounding HII and PDR regions
may - as noted by the authors - add uncertainty to the deduced extinction curve.
This is a particular problem in the infrared
where a contribution of free-free nebular emission may  artificially increase the derived $R_{V}$ value.
Using a large sample (154) of obscured OB stars, \cite{1995ApJS..101..335H},                              
concluded that even though the $R_{V}$ value may vary between 2.6 and 4.6,
the near-IR ($ \lambda \,> 0.9 \mu$m) extinction curve can be well-fitted to a power law with an exponent $-$1.73$\pm$0.04.
The variation in $R_{V}$ value indicates that
the light paths may partly pass through dark clouds, but this aspect is hard to quantify.
In an investigation of the Taurus cloud complex, \cite{2001ApJ...547..872W} observed 27 background,  
early type stars from the U to the K band, and most of these stars (23)
have only moderate extinction ($A_{V}\,<\,3.4$) and
the corresponding $R_{V}$ values average around 3.0, i.e. typical of the diffuse ISM.
The remaining four stars, with $A_V$ values in the range 3.6--5.7, have higher $R_{V}$ values,
indicating that these light paths probe denser regions with larger particles. \\
\\
In our view, there is a need for further studies of the extinction curve in deeper parts of dark clouds
and it is important, as far as possible, to include the UV/blue region
since grain growth will first affect the extinction curve at shorter wavelengths.
Thus, even though a "universal" power law may describe
the shape of the extinction curve well in the near-IR, the particle size distribution,
and thus the column density of dust mass, may be poorly determined.
The 2Mass all-sky survey (\cite{2006AJ....131.1163S}) has proven
extremely useful in providing extinction maps of dark cloud complexes,
and it is important to find out how well the conversion factor $N_{H_{2}}/E_{J-K_{s}}$ can be determined.
This is the main scope of the present investigation.
We focus on the well known dark globule B\,335, and based on multi-colour observations,
we apply different ways to determine the extinction and,
in particular, a new method that allows us
to both classify the background stars and determine the extinction.
For comparison, we also include observations of an early type star behind the Cha\,I cloud.\\

\section  {Observations and data reductions}
B335 (RA(2000)\,=\,294.25, DEC(2000) \,=ã7.57) and a reference  field (RA(2000) = 293.54, DEC(2000) = 7.62),
were observed using the NTT at La Silla during four
nights 2006-06-27--29. The reference field, hereafter called the {\it free field}
(\emph{B335ff} in the table) was selected to be at the
same galactic latitude as  B335 and free from cloud extinction.
The observations are summarized in Table \ref{obslog}.\\
\\
\begin{table}
\begin{minipage}[t]{\columnwidth}
\caption{Observation log.}
\label{obslog}
\renewcommand{\footnoterule}{}  
\begin{tabular}{llllll}
\hline \hline
object &camera &filter &centre &exp time&exp time\\
& & &$\lambda $ &ea image&total\\
& & &[nm]&[s]&[s]\\
B335 &EMMI-blue &U602  &354.0 &720 &24000\\
B335 &EMMI-red  &Bb605 &413.2 &720 &8500\\
B335 &EMMI-red  &g772  &508.9 &500 &4000\\
B335 &EMMI-red  &r773  &673.3 &300 &2000\\
B335 &EMMI-red  &I610  &798.5 &200 &1000\\
B335 &EMMI-red  &spec  &grism4 & & 6000\\
B335 &NOTCAM\footnote{summed image provided by M. G{\aa}lfalk} &Ks &2144. &900  &1000\\
B335ff &EMMI-blue &U602 &354.0 &720 &4000\\
B335ff &EMMI-red &Bb605 &413.2 &720 &3000\\
B335ff &EMMI-red &g772 &508.9 &500 &2000\\
B335ff &EMMI-red &r773 &673.3 &300 &1000\\
B335ff &EMMI-red &I610 &798.5 &200 &500\\
Cha035&EMMI-blue&U602 &354.0 &720 &7100\\
Cha035&EMMI-red&Bb605&413.2 &720 & 720\\
Cha035&EMMI-red&g772&508.9 &480 & 480\\
Cha035&EMMI-red&r773&673.3 &300 & 300\\
Cha035&EMMI-red&I610&798.5 &300 & 300\\
\hline
\end{tabular}
\end{minipage}
\end{table}
All object frames have been preceded and followed by
exposures of the standard star  SA111-1195  (\cite{1992AJ....104..340L})
in respective filter. The basic reductions (bias subtraction,
dark correction, flat-fielding, cosmic ray reduction) were carried
out using standard $IRAF$ routines. Then the $IRAF$ astrometric programs
are used for registering and co-adding the images using $Skyview$
facilities for astrometric data on stars common to all images of an
object. The individual frames were noise weighted by the
inverse of the errors of the stars in the middle magnitude range.
The star finding program $sextractor$ has then been used to
tabulate stars and positions. Photometry of the co-added images was
then  carried out using the $DAOPHOT$ photometry package.\\
\\
The Landolt equatorial standard star SA111-1925 was classified
with the help of the SED simplified method (see below) to be an A3V star.
Using stellar models and the filter characteristics
the colour correction from \cite{1992AJ....104..340L} filters    
(including the detector spectral response and the atmospheric transmission)
to the Eso NTT-EMMI Gunn and Bessel filter sets could be made.\\
\\
These data have been combined with data  from the Two Micron All Sky Survey (2MASS) Point
Source Catalog(PSC) (\cite{2006AJ....131.1163S}) thus extending the
stellar data with the $J-H-K_{s}$ measurements (where these exist) in the
B335 and the free field B335ff.\\
\\
The EMMI-images and the 2Mass data have been complemented by an image taken with the
NOTCAM  at the Nordic Telescope on La Palma using the $K_{s}$-filter (by M G{\aa}lfalk,
2005-06-30. (The image reduction is described in \cite{2007A&A...475..281G}).
The photometric calibration was based on the 2Mass field stars with low photometric errors.\\
\\
Finally $IRAC$ data from the {\it Spitzer} data archive have been used.
The {\it Spitzer} archived images were analyzed with the Mopex-software
package (\cite{2005ASPC..347...81M}, \cite{2005PASP..117.1113M},
\cite{2004PASP..116..842M}) to give the stellar magnitudes.\\
\\
All the stellar data have finally been compiled in tables for B335
and the free field containing object positions, magnitudes and
magnitude errors in all the relevant filters.\\
\\
\section{Results}
\subsection{The pair method}
There are a few stars in the B\,335 region for which we have spectroscopy.
Some of these have been classified as K7III stars.
They are obscured to varying degrees. One of them, No. 2, is more heavily obscured.
By comparing these stars, and normalizing to $E_{I-K_{s}}$,
we directly get the extinction curve (apart from the extrapolation
to zero wave-number i.e. $A_{K_{s}}/E_{I-K_{s}}$). Unfortunately,
the faintness of the obscured star in the U band results
in a large uncertainty regarding the the derived extinction in the UV.

\subsection{Statistical reddening}
\begin{figure}
   \centering
   \includegraphics[width=9cm]{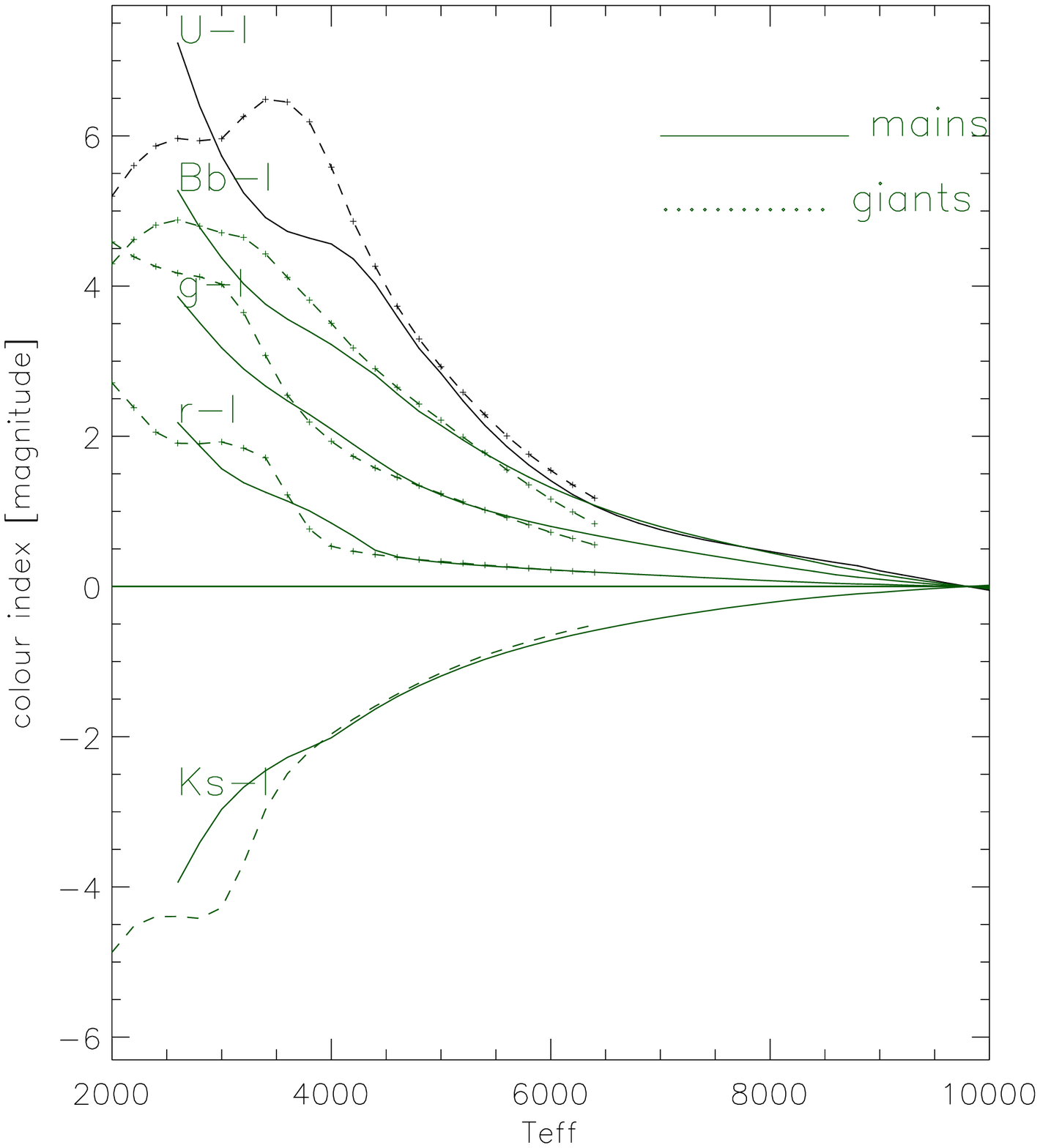}
      \caption{The colours as derived from the stellar atmospheric models
      from \cite{1999ApJ...525..871H}}
      \label{HauschildtColours}
   \end{figure}

\begin{figure}
   \centering
   \includegraphics[width=9cm]{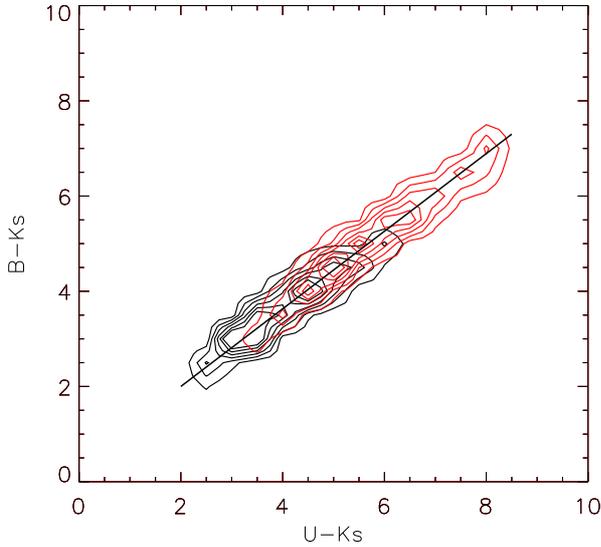}
      \caption{The contours represents the number density of colour indices for the stars in the B\,335 field
      and the reference field (black). The fitted line represents the direction of the reddening vector.  }
      \label{reddening}
   \end{figure}

Lacking any information on the intrinsic
SED:s (spectral energy distributions) of the stars, the traditional method is
to determine the {\it reddening vector} for the various colour index combinations.
The main problem in this approach is the wide spread of the intrinsic colours of the stars,
as is shown in  Fig \ref{HauschildtColours}.
For this reason it is useless for estimating the extinction towards individual anonymous stars.
However, it is a robust and simple method and worth looking closer at.
In Fig \ref{reddening} we show the
distribution of the colour indices $B-K_{s}$ versus $U-K_{s}$ for the two fields.
The expected intrinsic scatter is clearly seen in the free field,
but it is also clear that the contours
for the B\,335 field are both stretched and shifted due to the reddening.
The slope of the line, $E_{B-K_{s}}/E_{U-K_{s}}$, is well defined.\\


   \begin{figure}
   \centering
   \includegraphics[width=9cm]{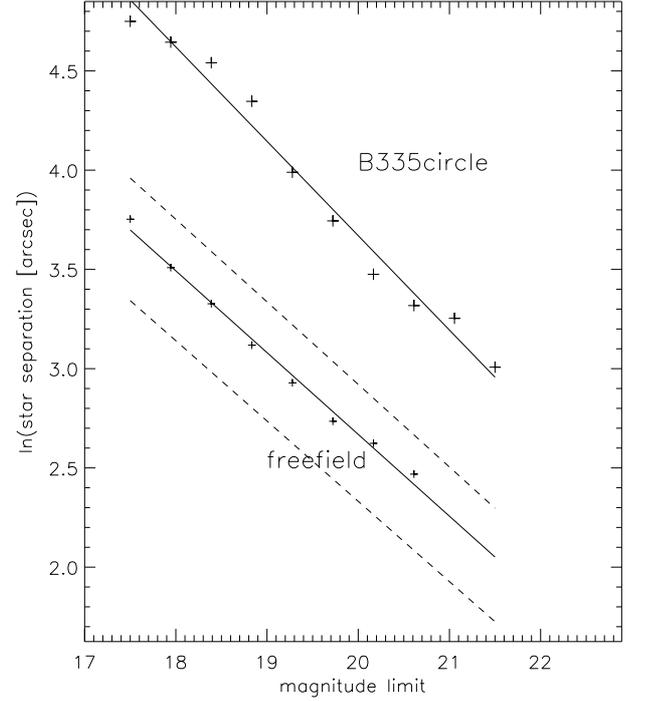}
      \caption{The median of the logarithmic projected separation
      (of the seven nearest) stars vs the U-magnitude limits.
      The dashed curves shows one standard deviation for the object in the B335circle.}
      \label{Cambresy}
   \end{figure}
\subsection{Star counting}
Star counting is a straightforward way of estimating the interstellar extinction.
The star count method goes back to \cite{Wolf23} and is based on the projected surface density of stars
in an obscured area compared to that of a region without (foreground) extinction.
The stars in magnitude intervals are counted in each cell of a grid of cells.
Later the method was improved to instead count the number of
stars up to a magnitude level. In this way the chosen grid cell size
is a compromise between spatial and statistical resolution.
\cite{1999A&A...345..965C} presented an improved method where the statistical resolution    
is fixed by measuring the local star density expressed as the projected distance
to the stars in the local surrounding. In the free field a relation of the mid (mean or median) distance $d_{free}$
of the neighbouring  stars up to a magnitude limit $m_{limit}$ gives a linear
relationship as shown in Fig \ref{Cambresy}.\\
\begin{equation}
log(d_{free}) = C_{linear}\cdot m_{limit}
\end{equation}
where $C_{linear}$ is a constant. This relation is linear in a range
determined by the completeness level of magnitudes.
For the reference as well as for the object field we have
the relation $m_{limit}\ =\ m_{0}\ -\ (1./C_{linear})\cdot log(d)$ see Fig \ref{Cambresy},
differing by different values for the constant $m_{0}$ in the reference and the object field,
namely the difference in the extinction between the two fields.\\
\\
   \begin{figure}
   \centering
   \includegraphics[width=9cm]{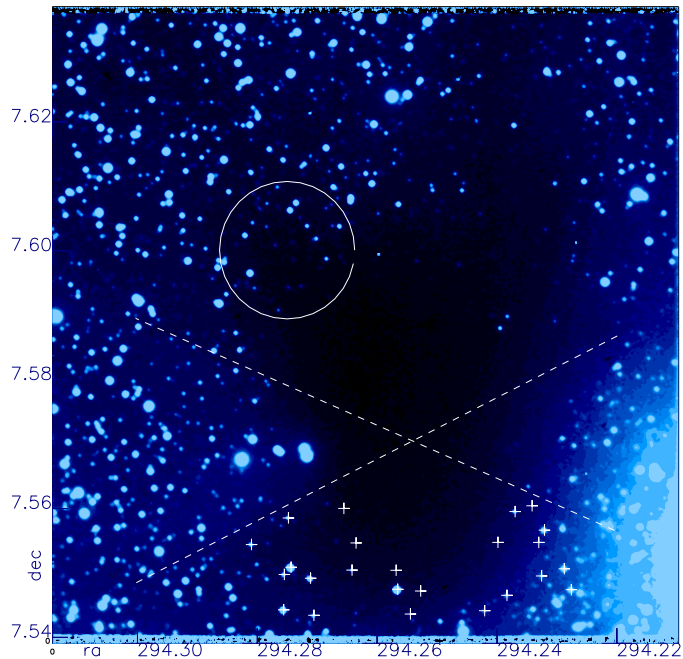}
      \caption{B335 in the U-filter with the circle shown in which
      extinction estimation by star counting has been performed in Fig \ref{Cambresy}.
      The dashed lines mark the cloud centre and outflow cones, + signs mark the stars
      that contribute in Fig \ref{BXtvsRadius} to the estimation of the cloud radius.}  
      \label {B335UsumBild}
   \end{figure}
%
The star count method is simple and straightforward. If the reference
field is representative for the star field behind the cloud, the
measurement gives the extinction of the foreground cloud.
In contrast to methods based on the colour of the stars, star counting gives the {\it total} extinction,
and the otherwise required extrapolation from the longest wavelength to infinity is avoided.
However, in practice the assumption on a smooth distribution of background stars is not valid and this limits the accuracy.
In addition, it gives a poor spatial resolution compared to methods based on reddening of the background stars.
Still, for comparison, we have applied the method of \cite{1999A&A...345..965C}        
and we have selected a sub-region in the B\,335 field, see Fig \ref{B335UsumBild}.   
The result is shown in  table \ref{Extinction_results}.

\subsection{The SED method}
Several multicolour systems (e. g. the Str{\"o}mgren four-colour \cite{1966ARA&A...4..433S},
or the Vilnius seven-colour photometric system \cite{1993ASNYN...4....7S})           
have been extensively used  to classify stars and quantify stellar properties. They are based
on measurements in well defined narrow-band filter sets and the correction for the interstellar extinction is carried out by {\it adopting} an extinction law.\\
\\
We propose a more general way to determine the intrinsic qualities of stars {\it as well as} the extinction curve,
based on multicolour measurements.
For a given star, the observed SED represents the combination of the intrinsic SED of the star,
the distance and the extinction along its light path.
Without any separate knowledge of the star and
without any assumption of the wavelength dependence of the extinction,
there is no way to determine the intrinsic SED of \emph{one} star \emph{and} the extinction.
However, if we consider two (or more) adjacent stars and assume
that their cloud extinction is the same, then it is possible to find both
the spectral class of the stars and the extinction of the intervening part of the cloud
(see Appendix A).\\
\\
There are practical limitations also for this method.
One is the assumption of equal extinction for adjacent stars,
which is risky in regions of strong gradients.
Another is the tendency for the reddening vector in a colour index diagram to be parallel to the locus of
the intrinsic colours of the stars as seen in Fig \ref{reddening}.
In other words, it may be difficult
to tell whether a star is red because it is intrinsically red or reddened by the extinction.
The degeneracy was resolved by choosing several combinations
of neighbouring stars and noting the consistency in the $T_{eff}$ determination.
It turns out that a broad spectral coverage is essential and
that it is indeed possible to find trustworthy solutions.
To represent the intrinsic SED:s of the stars we have used model atmospheres kindly provided by P. Hauschildt,
which cover the temperature range 2600\,K to 10000\,K at two surface gravities
$^{10}$log\,g equal to 0 and 4.5  representing the main sequence stars and
the giant stars respectively. In order to keep
the number of model SED:s at a manageable level, we only include one metalicity (solar).
The synthetic spectra cover the spectral range
$0.01 \mu \,<\lambda\,<\,100\,\mu$ at a high spectral resolution.
These model atmospheres are described by \cite{1999ApJ...525..871H}.   
The sensitivity function
(including filter transmission, detector sensitivity and atmospheric transmission)
was used to calculate the synthetic colours.\\
\\
   \begin{figure}
   \centering
   \includegraphics[width=9cm]{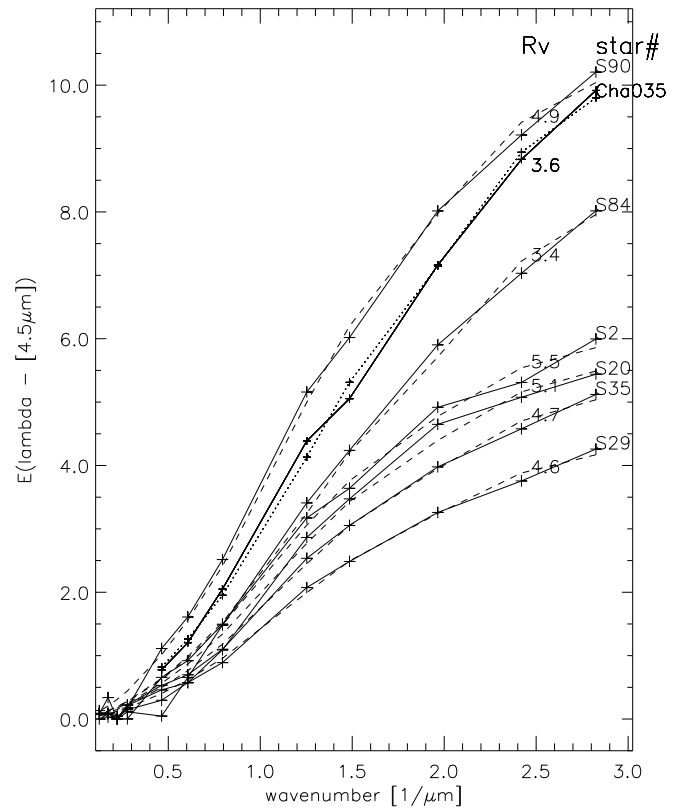}
      \caption{Extinction vs wavenumber $[1/\mu m]$ for several stars
      in the differently dense regions of the B335 cloud.
      The dashed curves are the $R_{V}$ fitted {\it CCM} curves.
      The curve marked Cha035 is a corresponding measurement of the extinction toward a star behind  the Chamaeleon I cloud.}
      \label{highXtinct}
   \end{figure}
Fig \ref{highXtinct} shows the extinction in the whole                   
wavelength range from 0.35 to 8.\,$\mu m$ for some stars in the
cloud border high-density area of B335. For each of these we fit a {\it CCM} curve.
We first note that {\it CCM} curves well describe the observed extinction curves and
that the $R_{V}$ values are in the range $R_{V}$ = 3.4--5.5 with an average of $R_{V}$ = 4.8.
A similar extinction curve constructed
from the measurements in the Cha\,I cloud (marked \emph{Cha035}) is also drawn in the same diagram
(together with its fitted {\it CCM} curve).\\
\\
   \begin{figure}
   \centering
   \includegraphics[width=9cm]{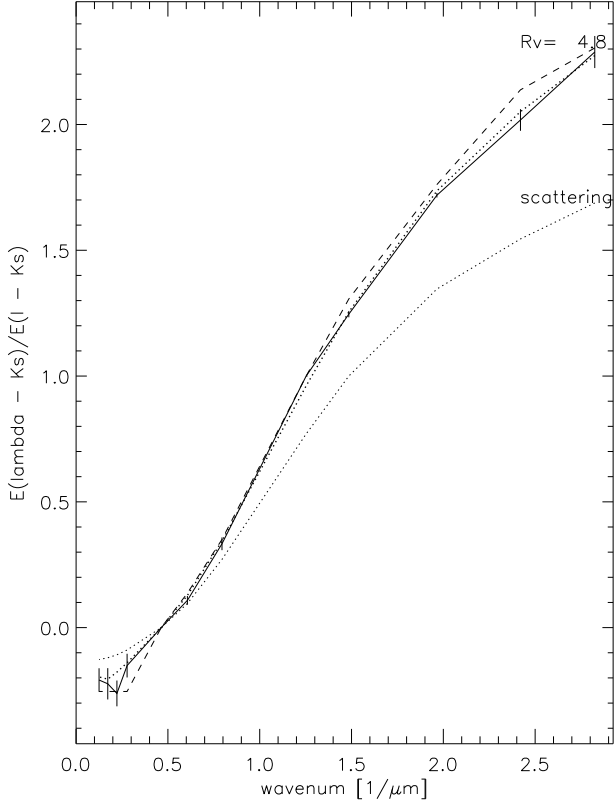}
      \caption{Extinction vs wavenumber $[1/\mu m]$ as the median for several stars
      (whose sight line extinctions are shown in \ref{highXtinct}).
      The {\it CCM} curve (dashed line) fitted to that extinction
      has a $R_{V}$ of 4.8. The grain distribution model shown
      in Fig \ref{WDgrainSizeD} gives a good fit to the observations (dotted line).
      The scattering part, labelled "scattering" shows the scattering part of the extinction.}
      \label{normzdXt}
   \end{figure}
The normalized and averaged extinction from B335 is
in Fig \ref{normzdXt} compared with a {\it CCM} curve with $R_{V}$ = 4.8.\\     
   \begin{figure}
   \centering
   \includegraphics[width=9cm]{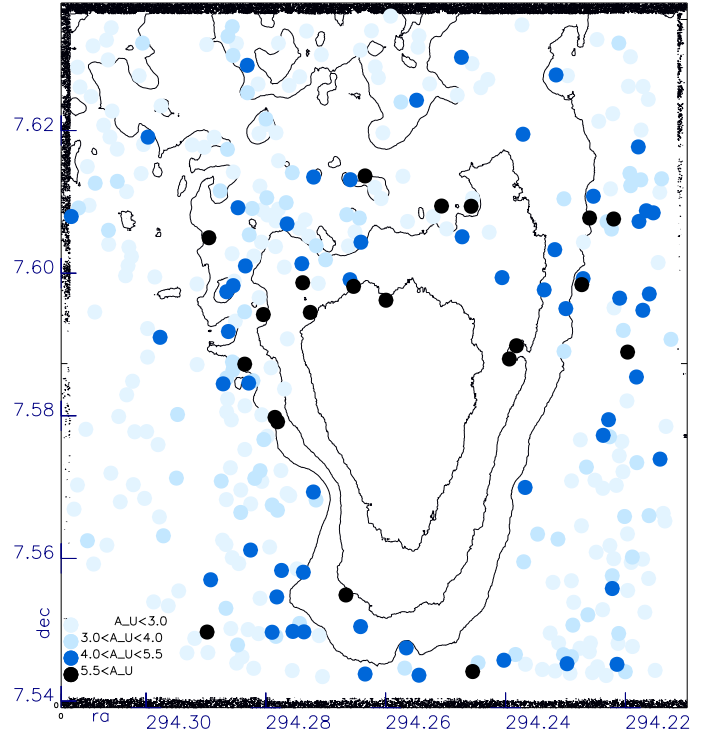}
      \caption{Extinction as calculated with the simplified SED method
      for each star (only stars with U and 2Mass Ks measurements are included).
      The contours mark the background flux in the U band due to the scattered interstellar radiation field.
      The step between contour lines is $\sim 5\cdot10^{-2}\,\mu Jy\,arcsec^{-2}$.}
      \label {B335XtperS}
   \end{figure}
\subsection{A simplified SED method}
\begin{figure}
   \centering
   \includegraphics[width=9cm]{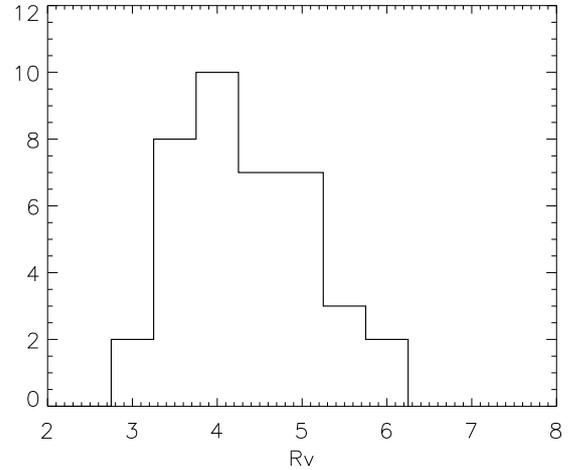}
      \caption{The $R_{V}$ distribution of the extinction
      towards the background stars with good photometry and with $E(I-K_{s}) > 3$.
      The median $R_{V}$value  is $\sim 4.8$.}
      \label{RvHistogram}
   \end{figure}

\begin{figure}
   \centering
   \includegraphics[width=9cm]{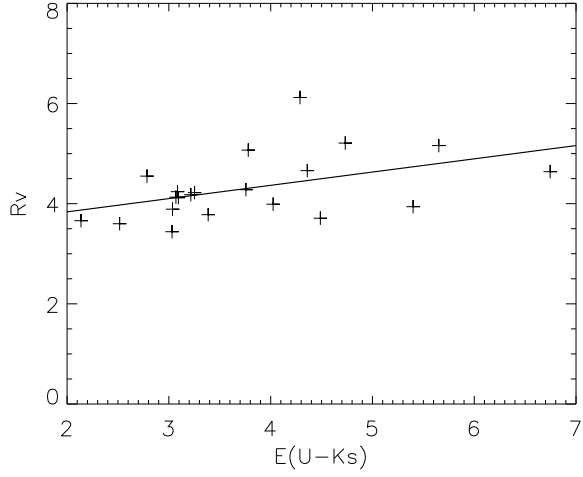}
      \caption{The $R_{V}$ for selected early type stars as a function
      of the excess $E_{U-K_{s}}$ showing a tendency for higher
      $R_{V}$ values at higher excesses.}
      \label{Rv_vs_E(U-Ks)}
   \end{figure}

All the methods tested above confirm that the extinction in the optical and NIR wavelength range
can accurately be described by the {\it CCM} curve.
This means that we can simplify the SED method by just leaving the $R_{V}$ value and
a colour excess (e.g.  $E_{I-K_{s}}$) as parameters and search for the best combination of R, E and model SED.
This can conveniently be done for each star resulting in an extinction map with high resolution,
as in Fig \ref{B335XtperS}.             
In this figure we compare the surface brightness
due to the scattered interstellar radiation field to the extinction of
the individual field stars. As expected, the stars with the highest extinction
are closest to the dark core of the globule.
In this figure we have included faint stars with relatively poor photometry.
Our optimization scheme allows the $R_V$ value to vary between 2 and 12 and,
surprisingly, we notice a large variation.
For faint stars with poor photometry as well as for stars with little extinction,
this can be explained as spurious, but how about the lines of sight
toward more obscured stars with accurate photometry - can we trust the deduced R$_V$ values?
If we restrict the sample to stars with $E(U-K_{s}) > 3$, $\sigma K_{s} < 0.3$
we get the $R_V$ distribution shown in Fig  \ref{RvHistogram}.
This sample includes late stars for which the model atmospheres probably are less reliable
and for which we expect larger departures from solar metallicity.
This means that the free parameter R$_V$ could to some extent compensate
for such non-perfect matching of the model colours to the true intrinsic colours.
If we restrict the sample further by only including stars with T$_{eff} > 7500$\,K,
for which we assume that the uncertainties are less we still find a scatter,
but interestingly also a trend; the R$_V$ value correlates to the extinction, see Fig  \ref{Rv_vs_E(U-Ks)}.\\
\\
We have applied the five methods in determining
the extinction curve towards B\,335 and in table \ref{Extinction_results} we summarize the results.
The different methods agree well.

\begin{table*}
\begin{minipage}[t]{\columnwidth}
\caption{Extinction descriptions}
\label{description}
\renewcommand{\footnoterule}{}  
\begin{tabular}{lll}
\hline \hline
reference & wavelength&description\footnote{
\begin{itemize} \item polynomials $f_{A}(x,\,N)$ and $f_{B}(x,\,N)$: x variable, N polynomial order;
\item C0 and C1 constants \item $R_{V}\,=\,A_{V}/E_{B-V}$\\
\end{itemize}}\\
& range $[\mu m]$ & $R_{V}\,=\,A_{V}/E_{B-V}$\\
\hline\\
\cite{1988duun.conf.....B} &$0.9\ \ <\ \lambda\ >\ 3.5$ &$A_{\lambda}/A_{V}\ \varpropto \lambda^{-\alpha}$\\        
\cite{1989ApJ...345..245C} &$0.35\ <\ \lambda\ <\ 0.9$ & $A_{\lambda}/A_{V}\ = \ f_{A}(1/\lambda,\ 7) + f_{B}(1/\lambda,\ 7)/R_{V}$\\  
 & $0.9\ \ <\ \lambda\ <\ 3.5$ &$A_{\lambda}/A_{V}\ = \ C0\cdot(1/\lambda)^{\alpha} + C1\cdot(1/\lambda)^{\alpha}/R_{V}$\\
\cite{1990ApJS...72..163F} & $0.11\ <\ \lambda\ <\ 0.3$ & 5-parameter model\\                                       
\cite{1999PASP..111...63F} &$\lambda\ >\ 0.11$& $R_{V}\ tabled\ A_{\lambda}/A_{V}$\\                                
\cite{2005ApJ...619..931I} &$1.25\ <\ \lambda\ <\ 5.$& $log(A_{\lambda}/A_{V})\ =\ f_{A}(log(\lambda),\ 2)$\\       
\hline
\end{tabular}
\end{minipage}
\end{table*}
\begin{table*}
\begin{minipage}[t]{\columnwidth}
\caption{Comparison of normalized extinctions.}
\label{Extinction_results}
\renewcommand{\footnoterule}{}  
\begin{tabular}{lllrrrrrrrrrrr}
\hline \hline\\
                       &   &           & \multicolumn{6}{l}{\textbf{EMMI}} & \multicolumn{3}{l}{\textbf{2Mass}} & \multicolumn{2}{l}{\textbf{Spitzer}}\\
\hline\\
\textbf{\textbf{band}} &   &           & U602  & Bb605  & g772  & V606  & r773  & I610  & J      & H      &$K_{s}$&$[3.6\mu]$&$[5.8\mu]$\\
$\lambda\ \ \ [nm]$    &   &           & 354.0 & 413.2  & 508.9 & 542.1 & 673.3 & 798.1 & 1258.5 & 1649.7 & 2157.2\\
$\Delta\lambda\ [nm]$  &   &           &  53.7 & 109.4  &  75.3 & 104.8 &  81.1 & 155.2 &  300.  &  300.  &  300. & 800. & 1500.\\
\hline\\
\textbf{source}& \textbf{medium} & \textbf{R$_{V}$} & \multicolumn{8}{l}{\textbf{E($[\lambda]$- Ks)/E(I - Ks)}}\\
\hline\\
this work\\
$\,\,\,$B335 SED-method & B335 & 4.8   & 2.29 & 2.02    & 1.72  &       & 1.25  & 1.00 & 0.33   & 0.11    & 0.00 &-0.15 &-0.22\\
$\,\,\,$B335 SED-stdv   &  &           & 0.06 & 0.04    & 0.01  &       & 0.01  &      & 0.03   & 0.02    &      & 0.05 & 0.06\\
$\,\,\,$B335 pair-method&  &           & 2.39 & 2.11    & 1.87  &       & 1.23  & 1.00 & 0.28   & 0.15    & 0.00 &-0.06 &-0.15\\
$\,\,\,$B335 pair-stdv  &  &           & 0.08 & 0.08    & 0.11  &       & 0.07  &      & 0.03   & 0.01    &      & 0.01 & 0.03\\
$\,\,\,$B335 statistical&  &           & 2.32 & 1.96    & 1.66  &       & 1.16  & 1.00 & 0.34   & 0.09    & 0.00\\
$\,\,\,$B335 stat-stdv  &  &           & 0.15 & 0.10    & 0.05  &       & 0.04  &      & 0.04   & 0.04\\
$\,\,\,$B335 star count &  &           & 1.9  & 1.7     & 1.5   &       & 1.3   & 1.00 & 0.00\\
$\,\,\,$B335 star-stdv  &  &           & 0.4  & 0.5     & 0.5   &       & 0.4   &      & 0.2\\
\hline\\
         &  \textbf{medium}& \textbf{R$_{V}$} & \multicolumn{8}{l}{\textbf{E($[\lambda]$- Ks)/E(I - Ks)}}\\
\hline
\cite{1989ApJ...345..245C} &     & 4.8  & 2.30 & 2.14    & 1.76  & 1.65  & 1.31  & 1.00 & 0.35 & 0.14     & 0.00 & -0.14 & -0.20\\ 
\cite{1985ApJ...288..618R} & dISM & 3.09 & 3.84 & 3.28    &       & 2.40  & 1.72  & 1.00 & 0.46 & 0.17     & 0.00\\           
\cite{1990ApJ...357..113M} & dISM & 3.0  & 3.91 & 3.22    &       & 2.37  & 1.70  & 1.00 & 0.44 & 0.16     & 0.00\\           
\cite{1995ApJS..101..335H} & dISM & 3.08 & 3.21 & 2.67    &       & 1.96  & 1.52  & 1.00 & 0.36 & 0.13     & 0.00\\           
\cite{1999PASP..111...63F} & dISM & 3.1  & 3.71 & 3.09    &       & 2.26  & 1.62  & 1.00 & 0.41 & 0.14     & 0.00\\           
\hline
\multicolumn{5}{l}{dISM \emph{stands for diffuse ISM}}\\
\end{tabular}
\end{minipage}
\end{table*}
   \begin{figure}
   \centering
   \includegraphics[width=9cm]{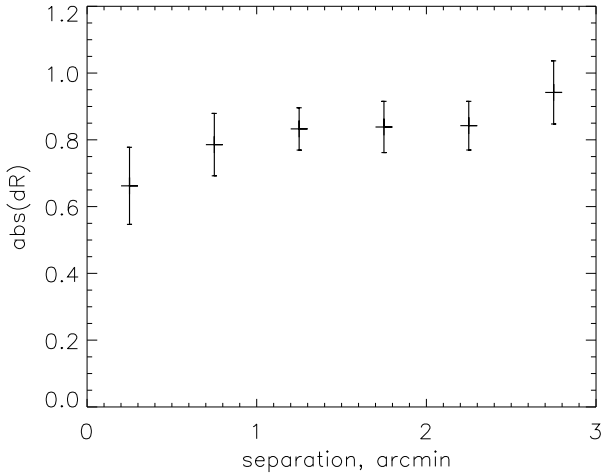}
      \caption{This figure, based on the same sub-sample as Fig \ref{RvHistogram},
      shows the statistical spatial correlation of the $R_{V}$ values.
      There is no significant tendency for adjacent stars to have similar $R_{V}$ values.}
      \label{dR_vs_ds}
   \end{figure}
\section{Interpretation}
\subsection{The extinction curve -- local variations?}
Our results show that {\it CCM} curves  well represent
the extinction in the  test cloud, B\,335 as well as for one location in the Cha\,I cloud.
The {\it simplified SED} method resulted in a distribution of R$_V$ values (Fig \ref{RvHistogram}),
and one may wonder whether there is any local variations,
apart from the tendency of a correlation between $R_{V}$ value and extinction \ref{Rv_vs_E(U-Ks)}.
In Fig \ref{dR_vs_ds} we show that there is no tendency for such local variation,
and we conclude that the scatter shown in Fig \ref{RvHistogram} is probably
due to non-perfect matching of the model colours to the real ones
as well as the observational uncertainties.
   \begin{figure}
   \centering
   \includegraphics[width=9cm]{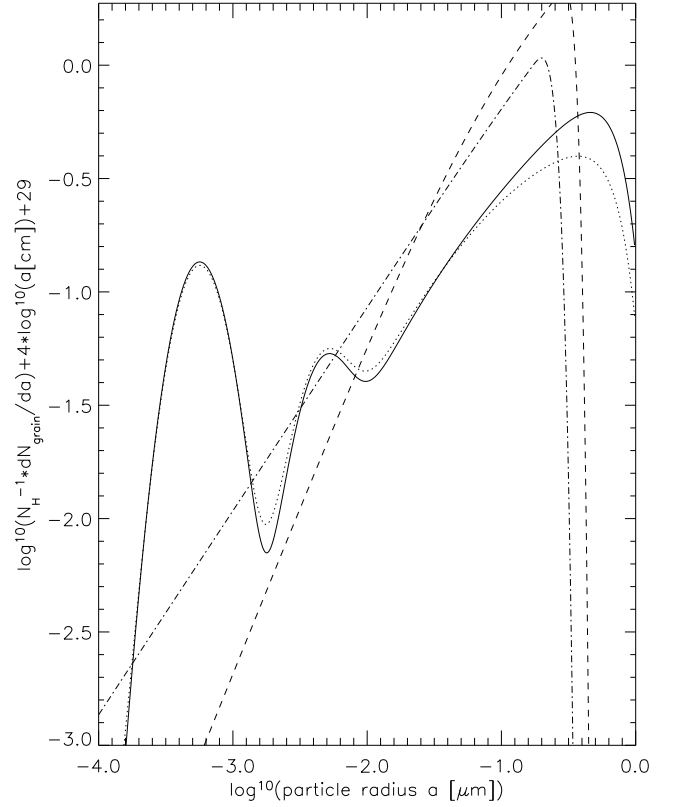}
      \caption{The grain size distribution for the average extinction as shown
      in Fig \ref{highXtinct} (full drawn line: graphite, dashed: silicates).
      As a comparison the \emph{WD2001} $R_{V}\,=\,3.1$ grain distribution
      dotted and dash-dotted. The model parameters used are shown in Tab \ref{WDpar}.
      ($dN_{grain}/da$ grain size distribution; \emph{a} grain particle radius; $N_{H}$ H-column density.)}
      \label{WDgrainSizeD}
   \end{figure}

\subsection{Grain size distribution}
We apply  the grain size distribution model constructed
by \cite{2001ApJ...548..296W} (hereafter \emph{WD2001}) and using Mie calculations (cf.  \cite{1983asls.book.....B})
we optimise the relative abundance of the different components to fit the observed extinction curve.
The model fit to the observed extinction is very good as shown in Fig. \ref{normzdXt}.
In Fig. \ref{WDgrainSizeD} we compare the resulting size distributions of the graphite
and the silicate grains to that for the diffuse ISM, and as expected the grains are larger in the molecular cloud.
Even though this general tendency for larger grains in the globule is a robust and expected result,
we cannot push the interpretation much further as the model fit
to the observed extinction includes  many parameters,
some of these not well constrained  by the observations.\\
\begin{table*}
\begin{minipage}[t]{\columnwidth}
\caption{Grain distribution parameters according to \emph{WD2001}}
\label{WDpar}
\renewcommand{\footnoterule}{}  
\begin{tabular}{lccccccccc}
\hline \hline\\
               & \multicolumn{9}{l}{\textbf{\emph{WD2001} grain model parameters}}\\
\hline
parameters                            & $b_{C}$ & $C_{g}$          & $C_{s}$              & $\alpha_{g}$ & $\alpha_{s}$ & $\beta_{g}$ & $\beta_{s}$ & $a_{t,g}$           & $a_{t.s}$\\
B335 (mean of several sightlines)     & 1.3 & $3.55\cdot 10^{-11}$ & $7.33\cdot 10^{-14}$ & -1.4         & -1.5         & -0.240      & -3.02       & $3.88\cdot 10^{-3}$ &  0.271\\
\emph{WD2001} table 1 $R_{V}\,=\,3.1$ & 4.0 & $2.9\cdot 10^{-11}$  & $1.3\cdot 10^{-13}$  & -1.8         & -2.1         & -0.132      & -0.114      &  8.98               &  0.169\\
\hline
\end{tabular}
\end{minipage}
\end{table*}
\\
   \begin{figure}
   \centering
   \includegraphics[width=9cm]{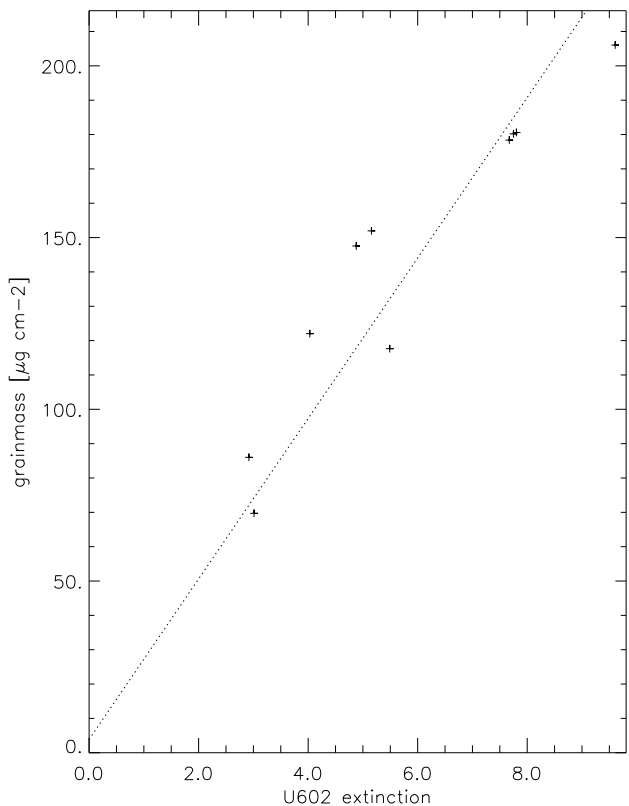}
      \caption{The total particle masses for the grain size distribution shown
      in Fig. \ref{WDgrainSizeD} as a function of the U-extinction.
      The slope of the curve corresponds to $23\,(\pm 3)\,\mu g\,cm^{-2}\ magnitude^{-1}$.}
      \label{WDmassXt}
   \end{figure}
\subsection{The column density}
In view of the possible spatial variations of the grain properties we
combine neighbouring stars in pairs, under the assumptions
\begin{itemize}
\item that the extinctions are represented by CCM functions
\item that they have similar grain size distribution and
\item that the extinction differences depend on a scale factor only.
\end{itemize}

Even though our model fits show variations in the carbon/silicate ratio,
the total amount of dust mass relates closely to the extinction,
and we find the following relation (Fig \ref{WDmassXt}):\\                           

grain mass = $23\, (\pm 2)\cdot A_{U}\,\mu g\,cm^{-2}$. \\
\\
We now turn to the average extinction curve ({\it SED method}) in table \ref{Extinction_results}.
Assuming that \emph{all} silicon is bound
in these grains as silicates and a 'Cosmic'
abundance of $[Si]\,= 3.63\cdot 10^{-5}\cdot [H]$ (\cite{1996ARA&A..34..279S})
we find the following relation:\\

$N_{H}\,=\,4.4\,(\pm\,0.5)\,\cdot 10^{21}\cdot E_{I-K_{s}}\,cm^{-2}$\\
\\
\noindent If we trust the model extrapolation to zero wave-number and interpolate
between the g- and r-filters we get the  relation:\\

$N_{H}\,=\,2.3\,(\pm\,0.2)\,\cdot10^{21}\cdot A_{V}\,cm^{-2}$\\
\\
\noindent for B335, which should be compared to the relations found in the literature
(e g \cite{1978ApJ...224..132B},                                
\cite{1995A&A...293..889P},                                     
\cite{1996Ap&SS.236..285R},                                     
\cite{2003A&A...408..581V})                                     
from UV Lyman $\alpha$ and X-ray analyses\\
$N_{H}\,=\,1.8 - 2.2 \cdot10^{21}\cdot A_{V}\,cm^{-2}$,\\
\\

   \begin{figure}
   \centering
   \includegraphics[width=9cm]{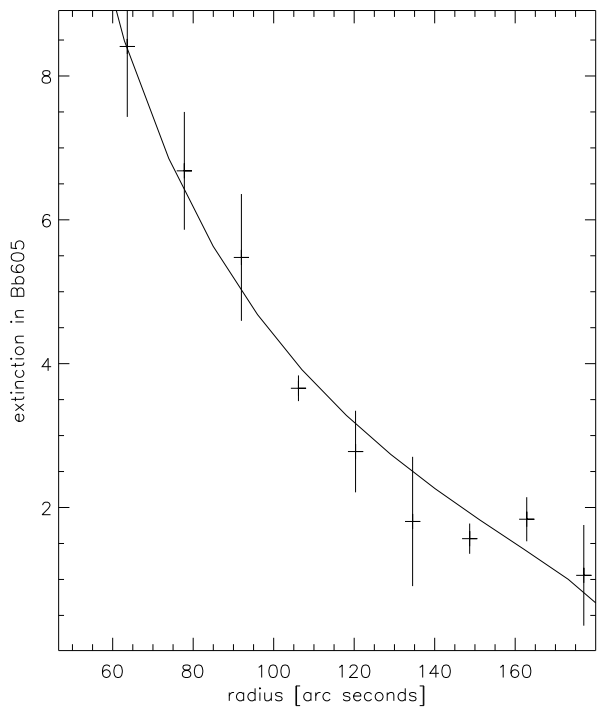}
      \caption{The B extinction vs radius [arcsec] from the centre at the protostar
      in the southern sector of the cloud (Fig \ref{B335UsumBild}). The curve marks the projected density
      for a gas sphere with $\rho\,\propto\,r^{-2}$ with an effective radius 190 [arcsec].}
      \label{BXtvsRadius}
   \end{figure}
\subsection{The mass of B\,335}
The young source in the middle of the globule has an East--West bipolar outflow with an opening angle of
$25^{\circ}\,\pm5$ (\cite{1988ApJ...334..196C} and \cite{1992ApJ...390L..85H}).             
However, the pre-stellar density profile in regions
away from the outflow (see Fig \ref{B335UsumBild}) has probably been roughly conserved.
We consider a sector in the south direction and plot the extinction as a function
of projected radius in Fig \ref{BXtvsRadius}.\\
\\
We can trace the extinction for $A_{B} < 10$,
which means outside a region at radii $> 1'$ from the cloud centre,
corresponding to a distance of 0.03 pc.
One of several models of the globule, that has been discussed
(\cite{1969MNRAS.145..271L}, \cite{1969MNRAS.144..425P}, \cite{1977ApJ...214..488S}),   
is the gas sphere evolved from the isothermal gravitationally compressed gas sphere
with an undisturbed outer shell and and a collapsing centre. In the undisturbed shell
the density varies as $\rho (r)\,=\,\kappa / r^{2}$ as described by \cite{1977ApJ...214..488S}.
The projected density can be fitted to the the extinction profile
deduced from sight-lines to the stars marked in the Fig \ref{B335UsumBild}
and shown in Fig \ref{BXtvsRadius}.
The effective cloud radius of the globule thus estimated is
of the order of $R_{cloud}\,\simeq\,190\,arcsec\,(\bigtriangleup R_{cloud}\,\pm \,\sim 20\,arcsec$).\\
\\
Thus assuming a gas sphere model with $\rho (r)\,\propto \,r^{-2}$ in line with the findings
by Harvey and coworkers (\cite{2001ApJ...563..903H}, \cite{2003ApJ...596..383H},
\cite{2003ApJ...583..809H}) we can estimate the mass $M_{B335}$ of the globule.\\
Given the cloud radius, the density gradient and the sound speed $a = \sqrt{k \cdot T/\mu}$ we get
\begin{equation}
M_{B335}\, =\,\frac{2\cdot a^{2} \cdot R_{cloud}}{G} \\
\end{equation}
where G is the gravitation constant.
From molecular line data \cite{1990ApJ...363..168Z} estimated the effective speed of sound      
$a$ to be $230\,m/s$ , which corresponds to a
kinetic temperature of $13^{o}K$. With the cloud model and the sound speed the mass $M_{B335}$ can thus
be estimated to be $2.2\,\pm\,0.2\,M_{\odot}$ (apart from the error in the estimate of the distance to the globule).\\
\\
The more direct way for handling this gas sphere-model
is to use the estimation of the measured silicate grain column density $\delta_{silicates}$
transformed into H-mass column density $\delta_{H}$ to get the globule mass $M$.
The protostar is assumed to be located in the centre of the cloud.                         
Thus with the impact radius b for a number of sightlines and
their H-mass column densities the simple calculations described in Appendix B
allow the globule mass to be estimated.

\begin{figure}
   \centering
      \includegraphics[width=9cm]{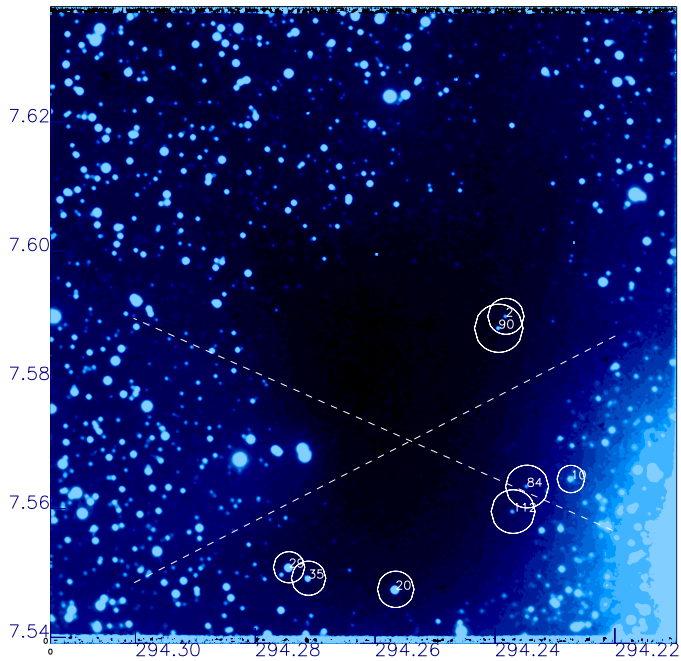}
      \caption{The map marks the sightlines for which the extinction measurements have been
      done in Fig \ref{highXtinct} and for which the globule mass has been estimated.
      The dashed lines mark the cloud centre and outflow cones.}
      \label{WDmap}
   \end{figure}

With the globule radius of $190\,\pm\,\sim 20\,arcsec$ the globule mass
for sightlines outside of the outflow cone
is found to be $2.5\,\pm 0.2\,M_{\odot}$.
The three sight lines through or near the outflow cone of the YSO in B335
(marked \#10, 84, 112 in the Fig \ref{WDmap}) result in lower mass estimates,
that average to $2.0\,\pm 0.1\,M_{\odot}$.
These total globular-mass estimates are in agreement with
the estimation of $2.2\,M_{\bigodot}$ done by \cite{2001ApJ...563..903H},
when corrected for the distance estimate (\cite{2009A&A...498..455O}.
That leaves a mass of less than 1.0 solar mass within
1 arcmin (the closest sight line in this study) from the centre star.\\
\\


   \begin{figure}
   \centering
   \includegraphics[width=9cm]{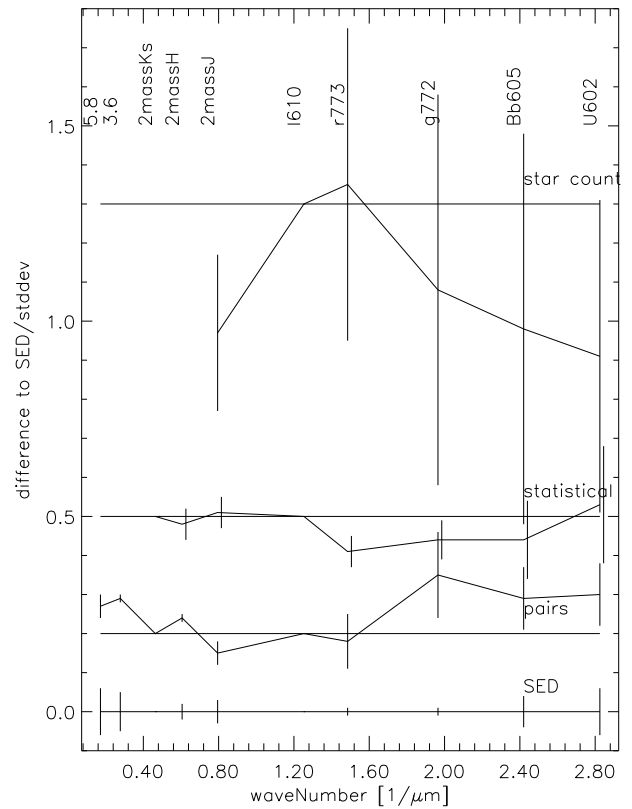}
      \caption{The difference between the extinctions from the different methods
      (from Tab \ref{Extinction_results}) and the SED-method with the standard deviations marked
      in the individual estimates as error bars.}
      \label{Xtstdv}
   \end{figure}
\section{Discussion}
As a next step it would be natural to investigate
a number of molecular clouds and the question is which method should be used.
We have used five different methods to determine the extinction curve in a molecular cloud.
Within the uncertainties, all methods give the same result
(see Tab \ref{Extinction_results} and Fig \ref{Xtstdv}).                    
There are advantages and disadvantages connected to
the different methods depending on the aim of the investigation.\\
\\
The {\it pair method} requires spectroscopic observations in addition to the multi-colour observations.
As the stars tend to be very faint in the optical region,
the spectroscopy should be carried out in the infrared.
The requirement of finding pairs
with identical spectral classes that in addition have significantly different extinction,
means in practice that quite an extensive spectrometry program should be planned for.
On the other hand, given the spectral class,
the intrinsic SED is known and the reddening curve can be determined to each star.
This is the method we used for the Cha\,
I 35 included as a comparison in Fig \ref{highXtinct}.
One potential problem is of course that the line of sight may
include some background extinction from the diffuse ISM.

The {\it statistical reddening} method is simple and robust but it suffers from the problem of separating
the large intrinsic scatter of the colour indices from the effect of reddening.
It can obviously not be used to gain
information on the extinction to individual  stars,
but it serves well as a complement to more detailed methods.

 The {\it star count} method is not well suited for determining the extinction curve.
It has several shortcomings and the only advantage,
that it in principle determines the {\it absolute} extinction
is not very useful as the accuracy is too poor.

 The {\it SED} method allows the determination of the extinction
in many sub-regions of the cloud and, like the previous methods,
it does not include any assumption on the functional shape of the extinction.
It is , however, computationally slow.

 The {\it simplified SED} method is based on our finding
that the extinction actually can be characterized by a {\it CCM} curve.
This may not necessarily be true in all clouds,
but it should suffice to first use e.g. the {\it statistical reddening}
method to check whether a {\it CCM} curve can be applied.

If so, the {\it simplified SED} method has
the advantage of defining the extinction curve towards each star.
It must of course be realized that e.g. non-resolved double stars,
having flatter SED:s than single stars, would cause spurious $R_{V}$ determinations.
This is probably part of the scatter seen in  Fig \ref{RvHistogram}.
Actually, it could for this reason be justified to exclude stars
with extreme $R_{V}$ values in constructing the extinction map of the cloud. \\

We have determined the extinction curve in the form $E_{[\lambda] - K_{s}}/E_{I - K_{s}}$,
and our model fit allows us to estimate the corresponding column density.
We have used  $E_{I\,-\,K_{s}}$ as the reference colour excess,
but in practice there is a number of different I band filters being used and for this reason,
and also because of the all-sky coverage of the 2Mass survey
it may be more useful to relate to $E_{J-K_{s}}$, even though the error bars are slightly larger.
The H-column density results are summarized in Tab \ref{Hcolumn} and compared
with some often cited literature values.\\

\begin{table*}
\caption{H-column density comparison.}
\label{Hcolumn}
\begin{tabular}{lllccccccccc}
\hline \hline\\
\textbf{source}          & \textbf{medium} & \textbf{method} & \multicolumn{4}{l}{\textbf{ratioes $N_{H}/A\cdot 10^{-21}\ cm^{-2}$}}\\
                         &                 &             & $N_{H}/E_{B-V}$     & $N_{H}/A_{V}$ & $N_{H}/E_{I-K_{s}}$  & $N_{H}/E_{J-K_{s}}$ & $N_{H}/E_{H-K_{s}}$ & $N_{H}/A_{J}$\\
\hline
this work                & B335        & multicolour     &                     & $2.3\,(\pm 0.2)$  & $4.4\,(\pm 0.5)$ & $15.3\,(\pm 1.9)$   & $47\,(\pm 8)$       & $7.5\,(\pm 0.8)$ \\
Bohlin et al, 1978       & diffuse ISM & Lyman $\alpha$  &   5.8               & 1.87\\
Bohlin et al, 1978       & $\rho$ Oph  & Lyman $\alpha$  &  15.8               \\
Predehl et Schmitt, 1995 & diffuse ISM & X-ray           &   5.3               & $1.79\,\pm0.03$ \\
Ryter, 1996              & diffuse ISM & Lyman $\alpha$ and X-ray & 6.8 \\
Vuong et al, 2003        & $\rho$ Oph  & X-ray           &                     &                    &                 &                      &                    &$5.6\,(\pm0.4)$\\
\cite{2007ApJ...669..493W}& Serpens    & X-ray           &                     &                    &                 &                      & 11.5 \\
\hline
\end{tabular}
\end{table*}

One problem  in going from the observed reddening curve (which relates to a colour index)
to the true extinction curve (which relates to a certain wavelength) is the extrapolation to zero wave-number.
Both the {\it CCM} curve and our dust model fittings provide this extrapolation,
but it would still be desirable to include measurements at longer wavelengths.
This will be presented in a forthcoming paper (Olofsson \& Olofsson, in preparation),
where we also will include ices in the measurements and the modelling.\\
\\
\section{Conclusions}
\begin{itemize}
  \item The extinction in the B\,335 globule follows a {{\it CCM}} curve with $4. < R_{V} < 6.$.
  \item A dust-to-extinction relation has been established: grain mass = 23.\,A$_{U}\,  \mu$g\,cm$^{-2}$.
  \item The relation between reddening  and hydrogen column density is $N_{H}\,=\,4.4\cdot\,10^{21}\,E_{I-K_{s}}\,cm^{-2}$
  \item Assuming a gas sphere with the outer shell modelled with a density profile as $\rho (r)\,\propto\,r^{-2}$,
   we find an effective globule radius of 190 arcsec. The mass of the B\,335 globule is estimated
   to be $M_{B335}\,=\,2.5\,M_{\bigodot}$
\end{itemize}
\begin{acknowledgements}
This publication makes use of data products from
the Two Micron All Sky Survey, which is a joint project of the
University of Massachusetts and the Infrared Processing and Analysis
Centre/California Institute of Technology, funded by the National
Aeronautics and Space Administration and the National Science
Foundation.\\
This work is based [in part] on archival data obtained with the
$Spitzer$ Space Telescope, which is operated by
the Jet Propulsion Laboratory, California Institute of Technology
under a contract with NASA.
\end{acknowledgements}

%% file: aa14174_appendix.tex
\begin{appendix}
\section{SED method} The analysis is based on models for stellar atmospheres
and for the interstellar medium extinction. The idea is that combining several
star in a group with neighbouring sightlines there is a part of the extinction common to all
in the group. This allows us to create an equation system containing
the colour indices $CI([\lambda_{i}]-[\lambda_{0}])$ from multi-filter measurements of
each star in the group. For a group of three we have for filter $[\lambda_{i}]$ and star\#1, 2 and 3
\begin{eqnarray*}
CI([\lambda_{i}]\ -\ [\lambda_{0}])_{1} & = & CI([\lambda_{i}]\ -\ [\lambda_{0}])_{0,1} + A_{\lambda_{i}}\ -\ A_{\lambda_{0}}\\
CI([\lambda_{i}]\ -\ [\lambda_{0}])_{2} & = & CI([\lambda_{i}]\ -\ [\lambda_{0}])_{0,2} + A_{\lambda_{i}}\ -\ A_{\lambda_{0}} + \Delta E_{2}\\
CI([\lambda_{i}]\ -\ [\lambda_{0}])_{3} & = & CI([\lambda_{i}]\ -\ [\lambda_{0}])_{0,3} + A_{\lambda_{i}}\ -\ A_{\lambda_{0}} + \Delta E_{3}\\
\end{eqnarray*}
where $CI([\lambda_{i}]-[\lambda_{0}])_{\ast}$ are the measured inputs,
while the $CI([\lambda_{i}]-[\lambda_{0}])_{0,\ast}$ are the parameterized stellar models
for the intrinsic colours with parameters
like effective temperature $T_{eff}$, surface gravity and metallicity.
Finally $A_{\lambda_{i}}\ -\ A_{\lambda_{0}}$ is the common excesses, namely that of star\#1.
The other stars may have the additional excesses $\Delta E_{\ast}$, caused by the ISM-extinction
between the background stars. As this also can be parameterized to follow e g a \emph{CCM}-function we have
a non-linear over-determined equation system, that can be solved by an optimizing technique
for $A_{\lambda_{i}}\ -\ A_{\lambda_{0}}$ and the stellar model parameters of the stars.\\
\\
In the application described the stellar models have been restricted to those with solar metallicity
and two surface gravities, one for the main sequence stars and one for giants.\\

\noindent Thus the number of parameters to be solved are
\begin{itemize}
\item $T_{eff}$ one effective temperature for each star.
\item $A_{\lambda_{i}}\,-\,A_{\lambda_{0}}$  one excess parameter for each filter measurement.
\item $\Delta E_{\lambda}$ : The interstellar extinction \emph{CCM}-characterization parameter $R_{V}$
for the determination of the $\Delta E_{\lambda}$'s. There is a proportionality constant
attached to each $\Delta E_{\lambda}$, as well, to determine all its $\lambda$'s.
\end{itemize}
\end{appendix}
\begin{appendix}
So, for three stars and eight filters the number of parameters are
3(Teff's) + 7(colour indices) + 1(\emph{CCM} $R_{V}$) + 2(proportionality constants)
equals 13 parameters (to be fitted to 21 measured colour indices).\\

\section{Mass determination from column densities.}
With an assumed model for the globule mass distribution and measurements of column densities it is
possible with simple calculations to get estimates of the globule mass. In this case we assume that
our column density measurements are made in the still undisturbed shell of the globule,
whose centre is known as well as the impact radius \emph{b}
for the column density measurement. Then with the following variables:
\[
\begin{array}{lp{0.8\linewidth}}
    R_{cloud}               & cloud radius\\
    \emph{b}                & impact radius from cloud centre\\
    m                       & calculated globule mass\\
    mol_{silicate}          & molecular weight of silicates = 172\\
    mol_{H}                 & molecular weight of H\,+\,He=\,1.25\\
    $[Si]$                  & abundance of Si\ \ $[Si]\ =\ 3.63\cdot10^{-5}\ \cdot\ [H]$\\
    \rho (r)                & density at radius r\\
    \delta_{silicate}(b)    & measured column density of silicates at b\\
    \delta_{H}(b)           & from $\delta_{silicate}$ calculated H-column density at b\\
    \end{array}
\]
The mass can be calculated in the following way:
\begin{eqnarray*}
    \rho (r)                  & = & \kappa\ /\ r^{2}\\       
    \delta_{silicate}(b)      & = & (2\ \kappa\ /\ b\ )\cdot \ \arccos(b\ /\ R_{cloud})\\
    \delta_{H}(b)             & = & \delta_{silicate}(b)\ \cdot mol_{H}/\ ([Si]\ \cdot\ mol_{silicate})\\
    \kappa                    & = & \delta_{H}(b) \cdot b/(2\cdot\ \arccos(b\ /\ R_{cloud}))\\
    m                         & = & 4\pi \ \int_{0}^{Rcloud} \rho\ \cdot\ r^{2}\ \cdot\ dr\\
    m                         & = & 4\pi\ \int_{0}^{Rcloud}  (\kappa /r^{2})\ \cdot\ r^{2}\ \cdot\  dr = 4\pi\ \cdot\ \kappa\ \cdot\ R_{cloud}\\
    m                         & = & 2\pi \cdot  \delta_{H}(b) \cdot b\cdot R_{cloud} / \arccos(b\ /\ R_{cloud})\\
\end{eqnarray*}
If more than one column density measurement at different impact radii \emph{b} are available
the parameter $R_{cloud}$ can be determined as well as the globule mass.\\
\end{appendix}